\documentclass[twocolumn]{aa}  

\usepackage{graphicx}
\usepackage{txfonts}
\usepackage{lipsum}
\usepackage{subcaption}         
\usepackage{lscape}             
\usepackage{placeins}           

\usepackage{natbib}
\bibpunct{(}{)}{;}{a}{}{,} 
\usepackage{comment}
\usepackage{float}
\usepackage{xcolor}
\usepackage[colorlinks=true,citecolor=blue,linkcolor=blue]{hyperref}
\usepackage{mhchem}
\newcommand{\Lxp}{L_X^\textit{f}}
\newcommand{\Lxa}{L_X^\textit{nf}}
\newcommand{\Txp}{T_X^\textit{f}}
\newcommand{\Txa}{T_X^\textit{nf}}
\newcommand{\Lsun}{L_{\odot}}
\newcommand{\Lfuv}{L_\mathrm{FUV}}
\newcommand{\Laccp}{L_{\mathrm{acc}}^{\mathrm{p}}}
\newcommand{\Lacca}{L_{\mathrm{acc}}^{\mathrm{a}}}
\newcommand{\Rsun}{R_{\odot}}

\newcommand{\micron}{\mu \mathrm{m}}
\newcommand \prodimo{\texttt{ProDiMo}}

\newcommand{\taurel}{\tau_{\rm{rel}}}
\newcommand{\spitzer}{\textit{Spitzer}}

\newcommand{\rev}[1]{{#1}}

\begin{document}

   \title{The Effect of X-Ray and Accretion Variability on Mid-Infrared Lines in the Young Disk-Bearing Binary DQ Tau
}



   \author{Bayron Portilla-Revelo \inst{1,2},
   Konstantin V. Getman \inst{1}, 
   Ágnes Kóspál \inst{3}, 
   Dmitry Semenov \inst{4,5},
   Vitaly Akimkin \inst{6},
   Lis Zwicky \inst{3,7},
   Jan Forbrich \inst{8},
   and Sierk van Terwisga \inst{9}}
   
\authorrunning{Bayron Portilla-Revelo et al.} 
\titlerunning{Effect of X-rays and Accretion Variability on Mid-Infrared Lines}
    
\institute{Department of Astronomy and Astrophysics, The Pennsylvania State University, 525 Davey Laboratory, University Park, PA 16802, USA \\
\email{bmp5924@psu.edu}
\and Center for Exoplanets and Habitable Worlds, Penn State University, 525 Davey Laboratory, 251 Pollock Road, University Park, PA, 16802, USA.
\and Konkoly Observatory, HUN-REN Research Centre for Astronomy and Earth Sciences, MTA Centre of Excellence, Konkoly-Thege Miklós út 15-17, 1121 Budapest, Hungary
\and Zentrum f\"{u}r Astronomie der Universit\"{a}t Heidelberg, Institut f\"{u}r Theoretische Astrophysik, Albert-Ueberle-Str. 2, 69120 Heidelberg, Germany
\and Max-Planck-Institut f\"{u}r Astronomie, K\"{o}nigstuhl 17, 69117 Heidelberg, Germany
\and Institute of Astronomy, Russian Academy of Sciences, 48 Pyatnitskaya St., Moscow 119017, Russia
\and Institute of Physics and Astronomy, ELTE Eötvös Loránd University, Pázmány Péter sétány 1/A, 1117 Budapest, Hungary
\and Centre for Astrophysics Research, University of Hertfordshire,
College Lane, Hatfield AL10 9AB, UK
\and Space Research Institute, Austrian Academy of Sciences, Schmiedlstrasse 6, 8042 Graz, Austria}

   \date{Accepted XXX}

 
  \abstract
  {A star's radiation field is the main energy source of a planet-forming disk. Stellar photons---from X-rays to optical---are reprocessed in the inner protoplanetary disk enabling mid-infrared (mid-IR) diagnostics of the disk's physical conditions. Most pre-main-sequence stars accrete at moderate rates and emit X-rays efficiently. Such activity is expected to drive stochastic variability in the stellar spectrum. It is not clear to what extent such variability affects the thermochemical structure of protoplanetary disks.}
   {We aim to examine the effect of routine accretion variability and X-ray flares on the thermochemical structure and mid-IR line emission spectra of protoplanetary disks.} 
  {We combine multiepoch, contemporaneous X-ray and NUV/optical observations of the DQ Tau binary with JWST/MIRI spectra of its circumbinary disk. We interpret the data and assess the effect of a varying stellar spectrum on the disk using a forward thermochemical model illustrative of the system.}
  {Synthetic mid-IR fluxes of \ce{CO}, \ce{CO2}, \ce{HCN}, and \ce{H2O} are systematically stronger at periastron than at apastron due to a fourfold increase in accretion luminosity. Variations in the integrated fluxes lie within a factor of two between epochs, in reasonable agreement with the JWST/MIRI spectra of DQ Tau. Routine stellar variability induces only modest changes in the disk's thermochemical structure, yet it can still shift the snowline outwards by $17$\% relative to its apastron position. Because the JWST observations did not coincide with X-ray flares, our flare results rely on simulations: moderate-intensity flares leave most mid-IR molecular emission unchanged; in contrast, X-ray flares strongly enhance the abundances and mid-IR luminosities of \ce{H I}, \ce{Ar II}, and \ce{Ne II} in the disk surface.}
   {Our results imply that the \spitzer/JWST line variability observed in non-outbursting systems is consistent with moderately variable stellar spectra driven by accretion and magnetospheric activity.}


   \maketitle
   \nolinenumbers

\section{Introduction}
\label{sec:intro}
Observations of protoplanetary disks with the \textit{Spitzer Space Telescope} and more recently with the \textit{James Webb Space Telescope} (JWST) have revealed infrared temporal variability. This variability is observed in both the continuum and the emission lines. JWST observations of the PDS 70 disk show a statistically significant flux enhancement around $7 \ \micron$ and $17 \ \micron$ compared to \spitzer, and similarly a $\sim 1$ year timescale variability is inferred from analysis of WISE time-series photometry \citep{Perotti2023,Jang2024}. An analogous behavior is seen in the optical and near-infrared, where strong spectroscopic and photometric variations occur on timescales of days to years \citep{Gaidos2024}. In the UX Tau A disk, an extreme "seesaw-like" variability results in a stark contrast between the substantial flux measured with \spitzer\ shortward $10 \ \micron$ and the nearly photospheric levels measured with JWST, $19$ years later \citep{Espaillat2024}. Similar "seesaw-like" patterns have also been reported in the T Cha \citep{Xie2025} and BP Tau \citep{Micolta2025} disks. Continuum mid-infrared (mid-IR) variability has \rev{also been observed in both single stars and binary systems. For example, multi-epoch surveys reported by \cite{Bary2009} show month- and year-long variability of the 10 $\micron$ feature in XZ Tau and DG Tau, the latter confirmed with spatially-resolved spectroscopy by \cite{Varga2017}}. In the binary system RW Aur A, \cite{Kurtovic2026} reports a nearly twofold decrease in the continuum flux beyond $10 \ \micron$ compared to \spitzer. Several ideas have been discussed in the literature to explain such continuum variability ranging from changes in the magnetic truncation radius, efficient dust grain growth, and variations in the inner disk geometry. 

Similarly, variability in emission lines has recently started to be detected. JWST/MIRI observations of the AS 209 disk show \ce{CO2} Q-branch fluxes nearly four times lower than those measured with \spitzer, and a twofold flux decline for other species commonly observed in the mid-IR such as \ce{OH} and \ce{H2O} \citep{Romero-Mirza2024}. For the VW Cha A disk, \cite{Kurtovic2026} find line fluxes two times higher compared to \spitzer\ for wavelengths beyond $12 \ \micron$. Variations in narrow features are also seen, as in the SZ Cha disk, where a variable \ce{[Ne III]}/\ce{[Ne II]} is reported \citep{Espaillat2023}. The origin of such emission variability for both broad and narrow features remains an open problem. 

An extreme example of line variability is seen in the EXor-type system EX Lup. Comparing \spitzer\ IRS spectra obtained before and during the 2008 outburst---during which a $50$x accretion luminosity boost was measured---\cite{Banzatti2012} reported a strong enhancement in the amplitudes of \ce{H2O} and \ce{OH} lines, whereas fluxes from \ce{C2H2, HCN} and \ce{CO2} disappeared. Follow up monitoring with JWST \citep{Kospal2023,Smith2025} show emission consistent with a quiescent, pre-outburst state, where organics are seen again, albeit cold water emission is still enhanced compared to other T Tauri disks.    

While not as extreme as in EXor-type stars, most classical T Tauris are expected to display variable episodes of accretion of moderate duration and amplitude (see \citealt{Fischer2023} for a review). Such "routine variability" induces optical changes of $\lesssim$ 1-2 mag, and of $\sim 0.13$ mag in the mid-IR \citep{Wolk2018}. 

Concurrently, T Tauri stars possess highly efficient
convection-driven dynamos operating throughout their fully con-
vective stellar volumes; these dynamos generate extensive coro-
nal active regions and intense magnetic-reconnection-driven X-
ray emission (e.g., \citealt{Feigelson2002, Preibisch2005, Getman2025}). \rev{The observed X-ray emission in T Tauri stars is
commonly interpreted as arising from a continuous distribution
of magnetic-reconnection flares spanning a broad range of energies. Frequent weak flares, unresolved in typical observations,
produce a quasi-continuous baseline (known as ``characteristic
emission''; e.g., \citealt{Gudel2003,Telleschi2005,Wolk2005}), while sufficiently energetic flares can be individually
detected as stochastic large-scale X-ray flares} \citep{Favata2005,Getman2008,Flaccomio2018}. While the X-ray 
baseline itself undergoes slow, multi-year
modulations, typically by factors $<2$ and possibly linked to
weak cyclical magnetic activity \citep{Getman2024}, \rev{this
quasi-continuous emission appears relatively stable on short
timescales but is frequently accompanied by individually
resolved energetic flares.} For young solar analogs, these powerful
flares---characterized by peak luminosities $L_{\rm X,pk}\sim
10^{30.5}$--$10^{34}$ erg s$^{-1}$ and total energies
$E_{\rm X}\sim10^{34}$--$10^{38}$ erg --- occur at estimated
rates of once every two days for $E_{\rm X}>10^{34}$ erg
events, and approximately three times per year for
$E_{\rm X}>10^{36}$ erg ``mega-flares'' \citep{Getman2021}. Such large X-ray flaring events are expected
to alter the physical and chemical structure of protoplanetary
disks (e.g., \citealt{Meijerink2012,Waggoner2022}).
     
Since the occurrence rate of X-ray flares is independent of the presence of a disk \citep{Getman2021}, accretion bursts and X-ray flares are generally understood as decoupled phenomena, although accretion can still contribute to a small portion of the soft X-ray emission \citep{Telleschi2007}. Coordinated multiwavelength observations of accretion and X-ray events are ideal to assess their combined effect on protoplanetary disks and their mid-IR observational signatures. However, as both physical mechanisms are inherently stochastic, such efforts are difficult to implement.    

Given its circumbinary configuration, an ideal test-bed to assess the effect of accretion and X-ray variability is the DQ Tau system. The inner binary pair are two nearly equal mass M0-type stars ($M_1+M_2=1.5 \ M_\odot$) orbiting their center of mass in a highly eccentric orbit ($e=0.56$) with a period of $15.80$ days \citep{Czekala2016,Fiorellino2022,Pouilly2023}. Periodic flux enhancements occur across DQ Tau's spectrum at orbital phases near periastron. In particular, X-ray and radio flares are triggered due to the stars' interacting magnetospheres \citep{Salter2010,Getman2011,Getman2023}, whereas accretion luminosity boosts are related to pulsed accretion driven by orbital eccentricity and disk perturbations \citep{Kospal2018,Muzerolle2019,Fiorellino2022}. Thus, in the case of eccentric binaries, both flaring and accretion events are expected to be modulated by the system's dynamics.

In this work, we integrate contemporaneous X-ray, NUV/optical, and JWST/MIRI observations of DQ Tau --- tracing magnetospheric activity, accretion, and disk spectral properties, respectively --- with thermochemical modeling via the \prodimo~ code \citep{Woitke2009,Kamp2010,Thi2011,Rab2018,Woitke2024} to assess the impact of X-ray and accretion variability on the physical structure and mid-infrared line emission of the disk. Section \ref{sec:methods} presents our dataset, describes the construction of the input stellar spectra for thermochemical modeling, and details the numerical star-plus-disk model implemented for DQ Tau. Section \ref{sec:results} presents our results, starting with the steady state response of the mid-IR spectra to a variable X-ray and UV/optical spectra, as well as that of the underlying disk thermochemical structure. The limitations of our model and the implications of our results --- specifically regarding the JWST-\spitzer\ line emission variability --- are discussed in Sect. \ref{sec:discussion}, followed by our conclusions in Sect. \ref{sec:conclusions}.

\section{Methods}
\label{sec:methods}

\subsection{\textit{Chandra}  and \textit{Swift} Observations} \label{subsec:chandra_and_swift_observations}
We conducted an observational campaign during winter--spring 2025 to study the influence of stellar optical, UV, and X-ray radiation on the physics and chemistry of DQ Tau's disk. The campaign involved JWST/MIRI, \textit{Chandra}/ACIS-I3, \textit{Swift}/XRT--UVOT, and ground-based facilities. \rev{The observations covered complementary wavelength regimes, including the mid-infrared with JWST/MIRI
($\sim5$--28~$\mu$m); X-rays with Chandra/ACIS-I (0.5--8~keV) and Swift/XRT (0.3--10~keV); and the optical and ultraviolet with Swift/UVOT (V, B, U, UVW1, UVM2, and UVW2 filters; $\sim$190--550~nm).} The JWST/MIRI spectra and ground-based data, along with the inferred effects of optical/UV accretion-driven radiation on the disk, are presented in \citet{Kospal2025}. Here, we describe the \textit{Chandra} and \textit{Swift} data reduction and analysis, including stellar X-ray luminosities and UVOT photometry, which are subsequently used in our ProDiMo modeling to assess X-ray-driven chemistry in the disk.

\textit{Chandra} monitored DQ Tau through 33 snapshots of 1.5~ks each, covering three periastron passages (January 28--31, February 13--16, and March 1--4, 2025) and one apastron passage (February 7--9, 2025; ObsIDs: 29680--29687, 29692--29699, 29704--29712, 29716--29723). Observations employed a one-eighth ACIS-I3 subarray \citep{Garmire2003} to mitigate possible pileup during X-ray flares. Data reduction used CIAO v4.17 \citep{Fruscione2006} and CALDB v4.12.0. The tools {\it chandra\_repro} and {\it reproject\_obs} were applied to reprocess and merge event images, and {\it srcflux} was used to measure count rates, apparent fluxes, and to generate spectra and response files.

\textit{Swift} observed DQ Tau during the same three periastron passages, producing 36 snapshots (ObsIDs: 00097800001--00097800026, 00097800028--00097800037) for a total exposure of 72~ks. The February 7--9 apastron was not observed due to the Moon's proximity. XRT operated in PC mode, while UVOT used the standard six-filter blue-weighted mode (0x30ed). XRT light curves and spectra were constructed using the Swift-XRT data product generator \citep{Evans2007,Evans2009} with HEASOFT v6.32. UVOT magnitudes (V:B:U:W1:M2:W2) were measured using {\it fappend} and {\it uvotmaghist} from HEASOFT v6.33.2 and CALDB UVOT data v20240201.

\begin{figure*}
\centering
\includegraphics[width=0.75\textwidth]{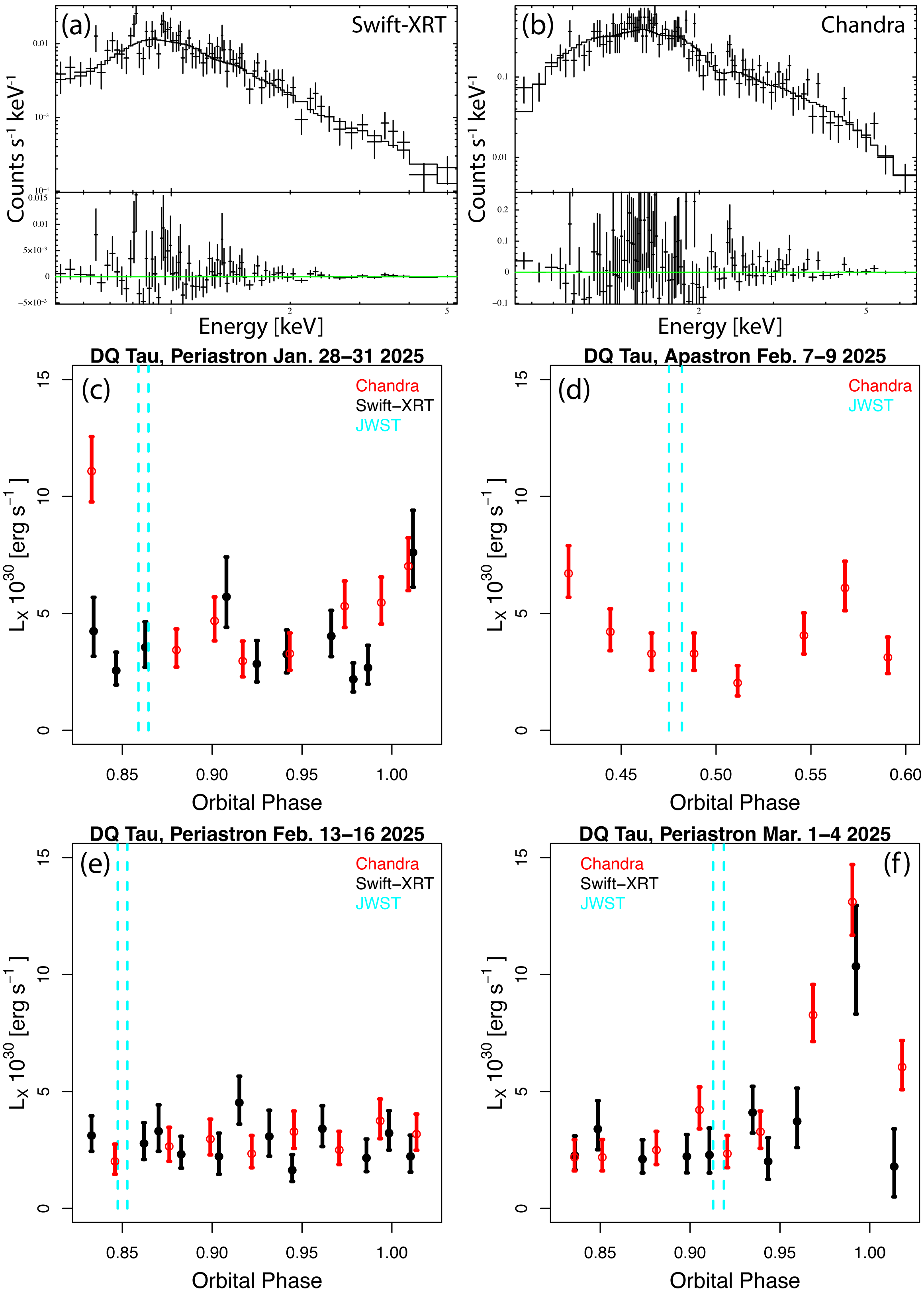}
\caption{X-ray spectra and light curves of DQ Tau obtained during our January--March 2025 campaign with \textit{Chandra} and \textit{Swift}. Panels (a-b): Time-averaged spectra combined across all \textit{Swift}-XRT (a) and \textit{Chandra} (b) observations. Data points with error bars represent the observed spectra; the best-fit models are shown as solid curves. The lower subpanels display the data-model residuals. Panels (c-f): X-ray light curves from \textit{Swift}-XRT (black points) and \textit{Chandra} (red points), with the timing of JWST/MIRI observations indicated by cyan dashed lines.}
\label{fig:chandra_swift_dqtau_2025}
\end{figure*}

Apparent X-ray fluxes were converted to intrinsic luminosities using stacked, time-averaged spectra (Figure~\ref{fig:chandra_swift_dqtau_2025}a,b). \textit{Chandra} and \textit{Swift}-XRT spectra were stacked separately across all relevant observations and fit in {\it XSPEC} \citep{Arnaud1996} with a two-temperature optically thin thermal plasma model subject to gas absorption ($tbabs \times (apec+apec)$). \rev{The coronal elemental abundances were fixed at 0.3 times the solar values in the {\tt apec} models, following a commonly adopted practice for young or magnetically active stars (e.g., \citealt{Imanishi2001,Getman2005,Gudel2007}). Given the limited signal-to-noise ratios and spectral resolution of our stacked Chandra and Swift spectra, adopting a single global metallicity of Z=0.3Z$_{odot}$ in the \texttt{apec} model provides an adequate approximation for deriving intrinsic X-ray luminosities, while more complex sub-solar elemental abundance patterns (e.g., implemented with \texttt{vapec}; \citealt{Gudel2007}) are not warranted for the present data.} 


Figures~\ref{fig:chandra_swift_dqtau_2025}(c--f) show the intrinsic X-ray luminosity as a function of orbital phase, with JWST/MIRI observation windows indicated. Several large X-ray flares above the baseline $L_{X,\rm base} = 2\times 10^{30}$~erg~s$^{-1}$ \citep{Getman2023} were detected; none coincided with the \textit{JWST}/MIRI windows. Time delays between apparent flare peaks and JWST observations were 0.5--0.8~days during January periastron and February apastron. Model--data comparisons, which are discussed in the following sections, suggest that DQ Tau's disk responds to stellar X-rays on timescales shorter than 0.5~days.

\subsection{Derivation of X-ray Spectral Properties}
\label{subsec:XraySpec}
We approximate the stellar radiation field by constructing a synthetic effective field that superimposes the contributions from both binary stellar components. Accordingly, the X-ray source is represented as a single central object with a total mass of $1.5 \ M_{\odot}$ \citep{Czekala2016,Fiorellino2022} and a radius of $R_\star=2.0 \ R_{\odot}$ \citep{Pouilly2023}.

Section~\ref{subsec:chandra_and_swift_observations} shows that several X-ray flares occurred across our four \textit{Chandra} epochs, although none were contemporaneous with the \textit{JWST} observations. During the \textit{JWST} observing windows, the X-ray emission remained close to DQ~Tau's baseline level of $L_X \sim 2\times10^{30}$~erg~s$^{-1}$. Regardless, one of the goals of the present modeling study is to predict how the disk chemistry responds to X-ray flares and to simulate the corresponding \textit{JWST}/MIRI spectra for future coordinated \textit{JWST}--X-ray observing campaigns. In the thermochemical modeling of DQ~Tau described below, we therefore conduct two trials: one adopting this baseline X-ray luminosity --- representative of the quiescent emission often \citep{Getman2022b,Getman2023}, but not always (Figure~\ref{fig:chandra_swift_dqtau_2025}d), seen near apastron --- and another adopting the higher luminosity characteristic of major periastron X-ray flares previously reported \citep{Getman2011,Getman2022b,Getman2023} and also detected in our current campaign (Figures~\ref{fig:chandra_swift_dqtau_2025}c,f), where $L_X$ reaches $\gtrsim  1.1\times10^{31}$~erg~s$^{-1}$. These luminosities are measured within the energy band $0.5$--$8$~keV. For simplicity, the modeling involving the baseline X-ray luminosity will be referred to as the non-flaring (``\textit{nf}'') case, while the modeling with the flare-level X-ray luminosity and higher stellar coronal plasma temperature will be referred to as the flaring (``\textit{f}'') case.    

To determine the plasma temperature ($T_X$ in [MK]) for the thermochemical modeling input, \rev{we fixed an average X-ray column density of $N_H = 1.3\times 10^{21}$~cm$^{-2}$}, and adopt a soft plasma component with $kT_1 = 0.7$~keV \citep{Getman2011,Getman2022b,Getman2023} assumed to be identical for both the non-flaring and flaring cases. This value is consistent with typical observations of young or older active stars and traces small-scale, fundamental stellar coronal structures \citep{Preibisch2005}. The temperature and emission measure of the hot plasma component differ between the non-flaring ("nf") and flaring ("f") scenarios. For the non-flaring baseline, we adopt $kT_2 = 2.6$~keV and $EM_2/EM_1 = 0.8$ \citep{Getman2023}. Although the current X-ray data lack sufficient counts and flare coverage for time-resolved spectroscopy to separate flare and baseline components (Fig.~\ref{fig:chandra_swift_dqtau_2025}), the duration and peak X-ray luminosity of the present periastron flares are comparable to those of the January 2010 event studied in detail by \citet{Getman2011}. For the flaring case, we adopt $kT_2 = 5.4$~keV and $EM_2/EM_1 = 5.4$ from \citet{Getman2011}. Over extended periods, the time-averaged values for $kT_2$ and $EM_2/EM_1$ fluctuate according to the observed frequency and intensity of flaring relative to the baseline emission. Specifically, across our entire observational campaign, we find mean values of $kT_2 \sim 2.5$~keV and $EM_2/EM_1 \sim 2$ (Sect. \ref{subsec:chandra_and_swift_observations}).

The characteristic plasma temperature for each modeled case is adopted as the emission measure-weighted mean of the soft and hard components:

\begin{equation}
    \langle k T_X \rangle =  \frac{kT_1+(EM_2/EM_1) kT_2}{1+EM_2/EM_1},
\end{equation}

\noindent from which we find, $\Txp=54$ MK a $\Txa=18$ MK.  

The total luminosity relates to the star's radiation field ($I_\nu$) via 

\begin{equation}
\label{eq:lumInu}
    L = 4\pi R_\star^2 \ \pi c  \int_{\lambda} I_\nu (\lambda)\ \lambda^{-2} \ d\lambda, 
\end{equation}

\noindent where $c$ is the speed of light. We model $I_\nu$ as a thermal bremsstrahlung spectrum with plasma temperature $T_X $ \citep{Glassgold1997}, $I_\nu \propto I_\nu^{\mathrm{Xray}} = a \cdot (h\nu)^{-1} \exp \big(-h\nu/\langle kT_X \rangle \big)$, where $a$ is a normalization constant satisfying Eq. \ref{eq:lumInu}. 

\rev{Although optically thin stellar coronal emission contains both continuum and numerous emission lines, as represented by the {\tt apec} spectral fits in Sect.~2.1 and previous X-ray studies of DQ Tau \citep{Getman2011,Getman2023}, the \prodimo \
calculations use a continuous incident X-ray spectral energy
distribution rather than an explicit coronal line spectrum.
We therefore adopt an analytically simple thermal
bremsstrahlung spectrum as a representation of the incident
X-ray radiation field, preserving the adopted plasma
temperature and total X-ray luminosity rather than the
detailed distribution of emission lines. To assess the validity
of this approximation, we compared intrinsic {\tt apec} and
thermal bremsstrahlung spectra with identical plasma
temperatures. {\it XSPEC} simulations show that the resulting
differences in the broadband spectral energy distribution are
small. For the ``f'' ($\langle kT_X\rangle \simeq 4.65$~keV) case,
the bremsstrahlung spectrum overestimates the luminosity
below 2.1~keV by $\sim6\%$ and underestimates the
luminosity above 2.1~keV by $\sim1\%$. For the ``nf''
($\langle kT_X\rangle \simeq 1.55$~keV) case, it underestimates
the soft-band luminosity by $\sim4\%$ and overestimates the
hard-band luminosity by $\sim4\%$. These small deviations are not expected to introduce any significant impact on the predictions of our thermochemical models.}

\begin{figure}
\centering
\includegraphics[width=\hsize]{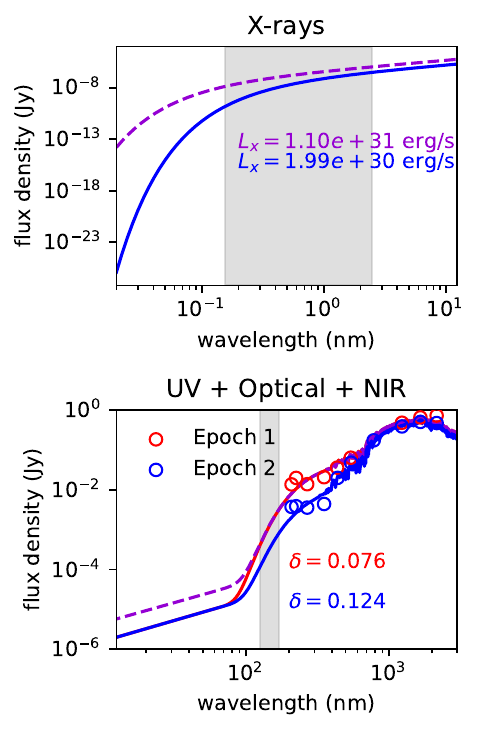}
\caption{Stellar flux densities at periastron and apastron. \textit{Top:} X-ray components in the interval between $62$ keV and $100$ eV. The blue curve indicates the X-ray baseline spectrum and dashed violet curve indicates the flaring X-ray spectrum. The shaded area covers the $0.5$ and $8$ keV interval where X-ray luminosities are calculated from the synthetic spectra and compared to the \textit{Chandra} observations. \textit{Bottom:} UV, optical and near-infrared components. The blue curve is the spectrum at apastron, the solid red curve indicates the spectrum at periastron for the X-ray non-flaring case, and the dashed violet curve represents the X-ray flaring case. \rev{The quantity $\delta$ represents the deviation of the FUV luminosity within the interval $125-170$ nm, indicated by the grey area, relative to the empirical relation by \cite{Yang2012}. Open circles indicate the photometry at both epochs.}}
\label{fig:starSpec}
\end{figure}

Adopting a distance of $d=196$ pc \citep{Gaia2021}, the X-ray flux densities are shown on the top panel in Fig. \ref{fig:starSpec}. As a sanity check, we integrate the radiation field over the shaded area which covers the energy range from $0.5$ to $8$ keV. Direct application of Eq. \ref{eq:lumInu} yields values of $\Lxa=2.0\times 10^{30} \ \mathrm{erg} \ \mathrm{s}^{-1}$ and $\Lxp=1.1\times 10^{31} \ \mathrm{erg} \ \mathrm{s}^{-1}$, which are equal to those measured from the \textit{Chandra} observations. 

\rev{Although powerful pre-main-sequence X-ray flares typically last from
several hours to a few days \citep{Getman2008,Getman2021b}, the
ProDiMo calculations are performed in a time-independent framework. We therefore approximate the effect of an X-ray flare by adopting a constant stellar X-ray luminosity and a harder X-ray spectrum representative of conditions near the observed flare peak. Consequently, our models do not simulate the temporal evolution of an X-ray flare, but instead represent
steady-state disk conditions under sustained flare-like X-ray
irradiation. While admittedly simplified, this approach
provides a useful first-order assessment of the impact of
powerful X-ray flares on disk thermochemistry.}

\subsection{Derivation of UV Spectral Properties}
\label{subsec:UVSpec}
The ultraviolet \rev{spectrum} is split into two components: the extreme ultraviolet (EUV: $13.6 \leq E < 100$ eV; i.e., $12.4 < \lambda \leq 91.2$ nm), and the far-, middle-, and near-ultraviolet, that we collectively referred to as UV ($5 \leq E < 13.6$ eV; i.e., $91.2 < \lambda \leq 400$ nm). 

EUV is modeled as a simple extrapolation of the X-ray spectra down to an energy of $13.6$ eV. Since the interstellar medium is highly optically thick to EUV photons, the $10-100$ nm portion of the spectra remains largely unconstrained, especially for pre-main-sequence stars. We acknowledge that this crude approximation likely underestimates the actual EUV flux, since X-ray observations primarily probe the hottest coronal plasma and do not strongly constrain the cooler transition-region and lower-coronal plasma that contributes substantially to the EUV emission (e.g., \citealt{Peacock2020} and references therein).  

The UV \rev{spectrum} is constrained by multiepoch SWIFT/UVOT photometry (sect. \ref{subsec:chandra_and_swift_observations}). The extinction-corrected fluxes taken with the \texttt{M2} ($\lambda \sim 225$ nm) and \texttt{W1} ($\lambda \sim 260$ nm) filters are displayed in the bottom panel in Fig. \ref{fig:starSpec}\footnote{These Swift/UVOT flux values are obtained from our current campaign data (Sect. \ref{subsec:chandra_and_swift_observations}). The high UVOT fluxes correspond to the $\Phi=0.91$ orbital phase during the January 28--31, 2025 periastron, when the combined optical/UV data indicate the peak of accretion activity \citep[see also Fig.~2 in][]{Kospal2025}. Conversely, the low UVOT fluxes were recorded at the $\Phi=0.84$ orbital phase at the onset of the February 13--16, 2025 periastron, when accretion activity was still at its global minimum \citep[][Fig.~2]{Kospal2025}.}. For each filter, flux variations of about a factor of five are seen between periastron and apastron. We interpret this NUV excess as being driven by the free-falling column of material accreted onto the stars \citep{Calvet1998,Ingleby2013}. 

At both periastron and apastron, the SWIFT/UVOT photometry is well explained by a $T=10^4$ K blackbody. The corresponding bolometric luminosities, equal to the accretion luminosities, are $\Laccp=0.4 \ \Lsun$ and $\Lacca=0.1 \ \Lsun$ \citep{Kospal2025}, respectively (see red and blue curves, bottom panel in Fig. \ref{fig:starSpec}). As a sanity check, we apply Eq. \ref{eq:lumInu} over the $125 \leq \lambda \leq 170$ nm range (shaded region in lower panel of Fig. \ref{fig:starSpec}) where \cite{Yang2012} found an empirical correlation between the far-ultraviolet luminosity and the accretion luminosity for classical T Tauris\footnote{For completeness, the \cite{Yang2012} relation is $\log\Big(\frac{\Lfuv}{L_\odot}\Big) = -1.670 + 0.836 \ \log\Big(\frac{L_\mathrm{acc}}{L_\odot}\Big)$.}. Within this wavelength range, we compute the relative error of the integrated $\Lfuv$ compared to the value predicted by \cite{Yang2012}, $\delta = |L_\mathrm{FUV}^{\mathrm{Yang}} - L_\mathrm{FUV}^{\mathrm{Integrated}}| \ /\  L_\mathrm{FUV}^{\mathrm{Yang}}$. The difference is $8\%$ at periastron and $12\%$ at apastron. 

In summary, we construct the total stellar spectra by adding the X-ray and UV contributions to the extinction-corrected stellar spectra (with no rim contribution) presented in \cite{Kospal2025}. As shown in Fig. \ref{fig:starSpec}, this provides an acceptable fit to the observed photometric data, from X-rays to the NIR. However, we note that although we were able to observationally constrain the NUV part of the stellar spectrum, the FUV part remains unconstrained. Since the FUV field is critical for thermochemical modeling (e.g., \citealt{Pegues2026}), we acknowledge this limitation in our methodology and advocate for future multi-epoch campaigns including FUV coverage.

\subsection{Thermochemical modeling}
We implement a thermochemical model for the DQ Tau circumbinary disk. Our model builds upon the dust radiative transfer model by \cite{Ballering2019}, constrained by photometric measurements of DQ Tau from the far-infrared to the millimeter. 

\cite{Ballering2019} made a few assumptions that are worth mentioning. First, they combined the radiation field of each star into a single central source, which aligns with our assumption of a central X-ray and UV source (sect. \ref{subsec:XraySpec}). Second, the spatial distribution for all grain sizes was assumed well-mixed with a gas component of constant gas-to-dust ratio equal to $100$. Last, dust extinction opacity considers only an absorption cross section and neglects extinction due to scattering off grains. While these assumptions are reasonable, we acknowledge their potential effect when modeling the thermochemical structure of a disk. In particular, dust settling can affect the temperature structure, and by extension the chemistry in a disk, since dust is the main source of opacity for UV wavelengths and beyond. We explore the effect of settling in Appendix \ref{subsec:effect-of-settling}. In this work, we do not attempt to modify any of the assumptions in \cite{Ballering2019} and do not aim at implementing a model that self-consistently explains dust and gas phase observations of DQ Tau at an absolute level. Due to this limitation, we focus only on reproducing relative flux measurements between periastron and apastron.     

Thermochemical modeling is carried out with the \prodimo \ code \citep{Woitke2009,Kamp2010,Thi2011,Woitke2016,Rab2018,Woitke2024}. \prodimo \ is a radiation thermochemical code that self-consistently solves for the continuum radiative transfer, the gas heating-cooling balance and the chemistry in a protoplanetary disk. The code assumes a two-dimensional geometry implemented on a cylindrical grid ($r,z$), where $r$ is the radial coordinate measured from the inner rim, and $z$ is the vertical coordinate measured from the disk midplane. Simulations are carried out with \prodimo \ in the \texttt{Master Branch}, revision \texttt{5dc72db0}. 

\begin{table}
\caption{\rev{Parameters of the thermochemical model illustrative of the DQ Tau circumbinary disk}}
\label{tab:model_properties}
\centering
\begin{tabular}{ll}
\hline\hline
Parameter & Value \\
\hline
Total Stellar Mass & $1.5\,\mathrm{M}_{\odot}$ \\
Effective Stellar Radius & $2 \, \Rsun$ \\
Mass accretion rate & $10^{-8}\,\mathrm{M}_{\odot}\,\mathrm{yr^{-1}}$ \\
\hline
Disk gas mass & $0.02\,\mathrm{M}_{\odot}$ \\
Gas-to-dust mass ratio & $100$ \\
Inner radius & $0.08\,\mathrm{au}$ \\
Tapering radius ($R_{\mathrm{tap}}$) & $104.7\,\mathrm{au}$ \\
Inclination & $22.7^{\circ}$ \\
Flaring exponent ($\beta$) & $1.11$ \\
Reference radial distance ($r_0$) & $100\,\mathrm{au}$ \\
Scale height at $r_0$ ($H_0$) & $19.2\,\mathrm{au}$ \\
\hline
Minimum grain size & $0.05\,\micron$ \\
Maximum grain size & $3630.8\,\micron$ \\
Grain size power index & 2.86 \\
Amorphous $\rm{Mg}_{0.7}\rm{Fe}_{0.3}\rm{SiO}_3$ (by volume)  & $60\%$ \\
Amorphous carbon  & $15\%$ \\
Porosity & $25\%$ \\
Carbon-to-oxygen elemental ratio & $0.457$ \\
\hline
\end{tabular}
\end{table}

We setup our simulations in parametric mode. The parameters that determine the disk's vertical structure are taken from \cite{Ballering2019}. For the radial distribution of gas we implement a similarity solution of a viscously evolving disk. At any given radius, the gas mass in a vertical column is found by rescaling the corresponding dust mass by a gas-to-dust ratio of  $100$. For the chemistry, we use the large chemical network from \cite{Kamp2017} with $236$ species and $3079$ reactions. \prodimo \ computes the gas temperature by balancing several heating and cooling mechanisms; this step requires knowing the level populations of the gas species included in the model. Level populations are computed assuming either LTE or non-LTE conditions, depending on the availability of collisional data. The levels of \ce{C2H2, HCN} and \ce{CO2} are all populated in a pseudo-nonLTE fashion; i.e., rotational levels are populated according to LTE whereas vibrational levels are populated according to non-LTE. A full non-LTE treatment is available for \ce{H2O} (collision data are taken from the LAMDA database, see \citealt{Schoier2005,vanderTak2020} and references therein); pure rotational \ce{OH} lines \citep{Offer1994,Rahmann1999,Tabone2021}; and for \ce{CO}, for which the custom molecular model by \cite{Thi2013} is used. Spectroscopic data are mostly taken from the HITRAN \citep{Gordon2022} and LAMDA databases.

Finally, synthetic mid-IR emission spectra are computed with an escape probability formalism. This approach ignores optical depth effects arising from overlapping lines. While considering such effects is important to reproduce fluxes on absolute scales, we emphasize that our focus is on the ratio of integrated line fluxes. 

\label{subsec:thermochemicalModelling}

\section{Results} 
\label{sec:results}

\subsection{The Effect of Accretion Variability And Baseline X-ray Emission on mid-IR Lines}
\label{subsec:effectOnSpec}

\begin{figure*}
\includegraphics[width=\hsize]{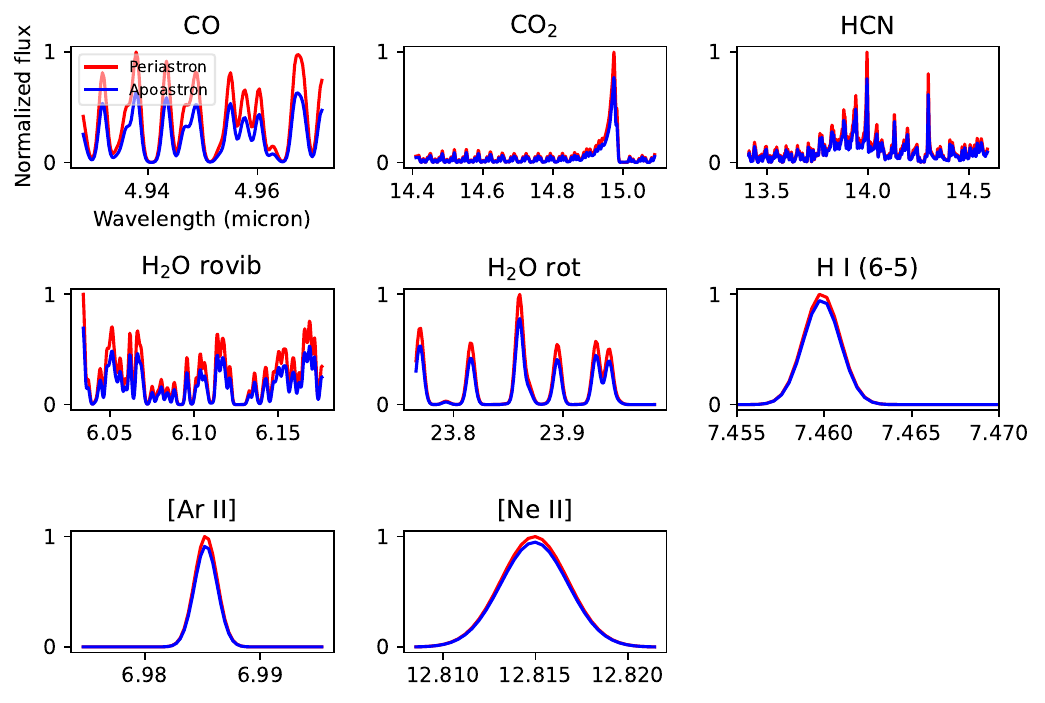}
\caption{Normalized synthetic mid-IR spectra. Each panel displays the flux contribution from a single molecular species. The curves correspond to two disk models: model 1 (red curves---simulation with an accretion burst at periastron) and model 2 (blue curves---simulation without an accretion burst at periastron). Both models utilize the "nf" X-ray variant, where only baseline X-ray emission is considered.}
\label{fig:spectra}
\end{figure*}

\begin{table*}[h]
\centering
 \caption[]{\label{tab:wavelength-ranges} Periastron-to-apastron integrated flux ratios for the observed JWST/MIRI data \citep{Kospal2025} and thermochemical disk models. The models utilize a consistent apastron baseline (no accretion burst and baseline X-ray levels) compared against two periastron scenarios: one including both an accretion burst and an X-ray flare ("f"), and another with an accretion burst but baseline X-ray levels ("nf"). Columns 1--2 list the molecular species and their associated spectral integration windows. Columns 3--5 provide the integrated flux ratios for the observed data and the two distinct model configurations. \rev{The observed ratio correspond to the 2025-Jan-29 periastron to the 2025-Feb-08 apastron.}}
\begin{tabular}{lcccc}
 \hline \hline
 Species & Wavelength Range & Observed Flux Ratio & \multicolumn{2}{c}{Modeled Flux Ratio} \\
 & & & ($\Lxa,\Txa$) & ($\Lxp,\Txp$) \\
 \\ \hline
\ce{CO}   & $4.925-4.975$ & $1.51 \pm 0.09$ & $1.58$ & $1.57$ \\
\ce{CO2}  & $14.40-15.10$ & $1.24 \pm 0.33$ & $1.33$ & $1.38$ \\
\ce{HCN}  & $13.40-14.60$ & $1.37 \pm 0.45$ & $1.22$ & $1.53$ \\
\ce{H2O} rovib  & $6.030-6.180$ & $1.28 \pm 0.19$ & $1.44$ & $1.38$\\
\ce{H2O} rot   & $23.75-24.00$ & $0.86 \pm 0.13$ & $1.30$ & $1.19$\\
\ce{H I (6-5)}   & $7.456-7.464$ & $5.65 \pm 0.36$ & $1.06$ & $2.66$ \\
\ce{[Ar II]}   & $6.982-6.989$ & $1.16 \pm 0.14$ & $1.10$ & $3.25$ \\
\ce{[Ne II]}   & $12.806-12.824$ & $1.15 \pm 0.05$ & $1.05$ & $2.85$ \\
\ce{H I (10-7)}   & $8.755-8.765$ & $3.37 \pm 0.43$ & $1.06$ & $2.84$ \\
\ce{H I (7-6)}   & $12.364-12.380$ & $3.89 \pm 0.46$ & $1.06$ & $2.75$\\
\ce{H I (8-7)}   & $19.044-19.080$ & $1.57 \pm 0.39 $ & $1.06$ & $2.82$\\
\hline
\end{tabular}
\end{table*}

We have run three thermochemical models of the DQ Tau disk, representing different stellar configurations tied to the system's orbital phases (Sects. \ref{subsec:XraySpec} and \ref{subsec:UVSpec}). These settings reflect the fact that our JWST/MIRI observing windows occurred during periods without large X-ray flares. The models are defined as follows:

\begin{itemize}
    \item \textbf{Model 1 (Periastron; ``nf''):} Includes an accretion burst but assumes baseline X-ray emission levels. 
    \item \textbf{Model 2 (Apastron):} Represents the quiescent state with no accretion burst and baseline X-ray emission.
    \item \textbf{Model 3 (Periastron; ``f''):} Includes both an accretion burst and a large X-ray flare.
\end{itemize}

We focus here on the comparison between the first two models and the JWST/MIRI data. Section \ref{subsec:isoXrays} is dedicated to detailing the results from the third model.

Synthetic emission spectra are generated for each species detected in the MIRI spectrum of DQ Tau (\ce{CO, \ CO2, \ HCN, \ H2O, \ H I, \ Ne II} and \ce{Ar II}) at each epoch. Figure \ref{fig:spectra} compares the modeled spectra for those species between the periastron (accretion burst-on; red) and apastron (accretion burst-off; blue) cases, both assuming the ``nf'' X-ray baseline. Fourth column in Table~\ref{tab:wavelength-ranges} lists the predicted periastron-to-apastron ratios of line fluxes integrated over the wavelength ranges indicated in the second column. 

The synthetic fluxes from atomic species at the disk surface remain essentially unaffected. \ce{[Ar II]} is the most responsive line showing only a 10\% enhancement relative to its apastron flux. Comparable variations are inferred for the \ce{H I} and \ce{[Ne II]} lines (see "nf" column in Table~\ref{tab:wavelength-ranges}). 

The signals from all molecular species are predicted to be stronger at periastron. The most reactive species is \ce{CO}, which shows a $58\%$ increase in the wavelength-integrated line flux, followed by the rovibrational \ce{H2O} component with $44\%$. The least reactive species is \ce{HCN} which exhibits only a 22\% flux increase at periastron. Finally, flux enhancements for \ce{CO2} and the pure rotational \ce{H2O} component are around $30$\%.

Overall, our models predict periastron-to-apastron flux ratios ranging from $1.1$ to $1.6$ for the lines detected in DQ Tau's mid-IR spectrum. Such predicted flux variations are a consequence of enhanced gas temperatures in the inner disk (see. Sect. \ref{subsec:convergedStruct}) driven by variable accretion between periastron and apastron passages.

\subsection{Application: The Case of DQ Tau}
\label{subsec:comparetoDQTau}
We compare the synthetic flux ratios with JWST/MIRI data from DQ Tau. As a first step, we subtract the best fit slab models from \cite{Kospal2025} from the continuum-subtracted observations to minimize contamination from overlapping lines within the same wavelength ranges used in Sect. \ref{subsec:effectOnSpec} (see Table \ref{tab:wavelength-ranges}). Within the ranges for \ce{CO2} ($14.40-15.10 \ \micron$) and \ce{HCN} ($13.40-14.60 \ \micron$), we subtract the best fit slab model for each species because their spectral lines overlap with one another. Additionally, we also correct for water contamination by subtracting the water best fit model in the range $13.4-15.1 \ \micron$. For the \ce{[Ar II]} line, we correct for water contamination within $6.9-7.3 \ \micron$ range. While the $P-$ branch of the \ce{CO} ro-vibrational spectrum is expected to be contaminated by water lines, this is mostly a concern beyond $5\ \micron$. Shortward this threshold, the emission is dominated by the $v=1-0,\ 2-1,\ 3-2$ transitions of \ce{CO}, and thus we consider water contamination to be negligible within our range of interest. All the hydrogen lines listed in Table \ref{tab:wavelength-ranges} are corrected for water contamination, except for the H I (10-7) line that we assume uncontaminated. A very small water contribution is also removed around the \ce{[Ne II]} transition. We note that the best fit slab models are separately determined for each epoch of interest: the periastron passage on 2025-Jan-29, and the apastron passage on 2025-Feb-8. 

\begin{figure*}
\includegraphics[width=\hsize]{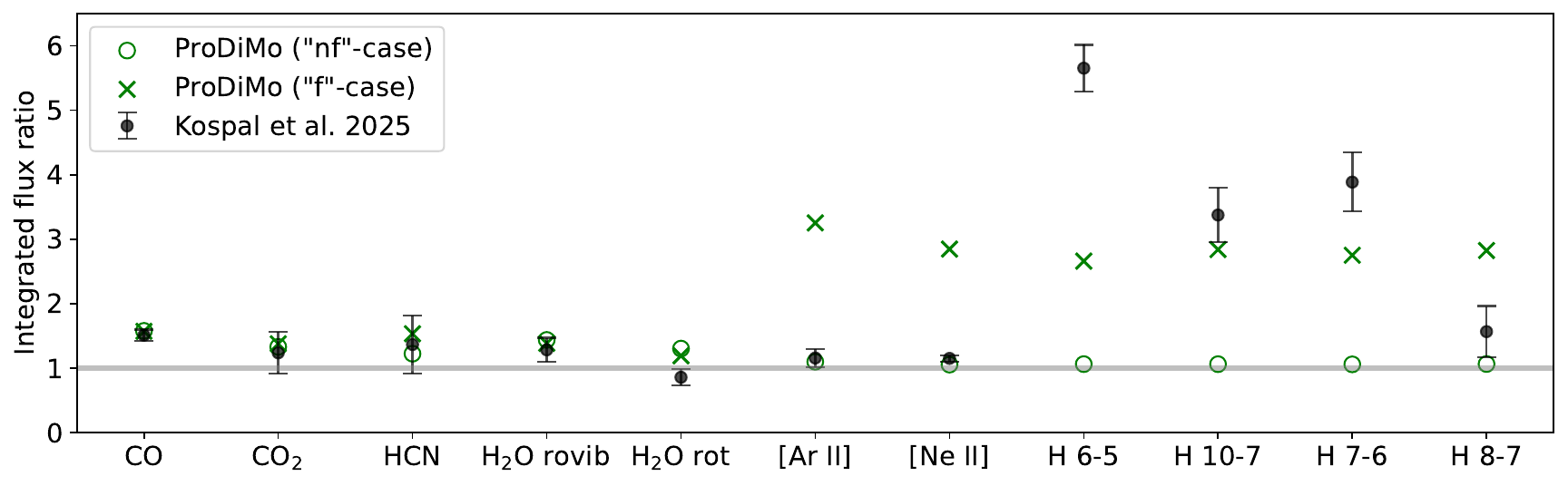}
\caption{Periastron-to-apastron integrated flux ratios. Black dots represent the ratios measured from the observed JWST/MIRI spectra \citep{Kospal2025}. The green $\circ$ symbols indicate ratios retrieved from \prodimo\ models corresponding to the ``nf'' scenario (maintaining baseline X-ray levels at both periastron and apastron), whereas green $\times$ symbols indicate the ``f'' scenario (including an X-ray flare at periastron). Points above the \rev{gray} horizontal line indicate a stronger signal at periastron than at apastron, while points below the line represent the reverse.}
\label{fig:fluxRatios}
\end{figure*}

We apply the analysis described in Sect. \ref{subsec:effectOnSpec} to the slab-corrected JWST/MIRI spectra: this is, we integrate the line fluxes over the intervals defined in Table \ref{tab:wavelength-ranges} for each epoch and calculate their respective ratios. \rev{We remind the reader that the periastron values correspond to the 2025-Jan-29 passage.} The results are presented in Fig. \ref{fig:fluxRatios} and in the third column of Table \ref{tab:wavelength-ranges}, where error bars represent the propagated 1$\sigma$ uncertainties derived from the total, non-continuum-subtracted observed spectra (black points). Additionally, Fig. \ref{fig:fluxRatios} displays the synthetic ratios from the two considered scenarios for periastron: assuming baseline X-ray levels (Sect. \ref{subsec:effectOnSpec}) and X-ray flare conditions. Since our JWST/MIRI observations were conducted during periods that missed DQ Tau's large X-ray flares (Fig.~\ref{fig:chandra_swift_dqtau_2025}), we defer the discussion of the model outputs including X-ray flares to Sect. \ref{subsec:isoXrays}.

The observed integrated flux ratios for the hydrogen lines are systematically larger than unity, indicating higher fluxes at periastron. H I (6-5) is the most responsive line, with an observed enhancement factor of 5.7, followed by H I (7-6), H I (10-7), and H I (8-7) with factors of 3.9, 3.4, and 1.6, respectively. This contrasts with the "nf" simulations, where enhancements do not exceed 10\%. As these hydrogen lines are established accretion tracers \citep{Rigliaco2015,Tofflemire2025,Shridharan2025} and our disk models do not account for emission from accretion columns, this large discrepancy between observed and synthetic ratios reaffirms an origin within the columns of accreting material. Conversely, the mean observed fluxes for \ce{[Ar II]} and \ce{[Ne II]} are only marginally higher at periastron and are well matched by the synthetic fluxes, which remain largely insensitive to changes in accretion luminosity and baseline X-ray activity. 

Regarding molecular species, our modeled periastron-to-apastron flux ratios for \ce{CO}, \ce{CO2}, \ce{HCN}, and \ce{H2O} rovibrational fall within the 1$\sigma$ confidence intervals of the observations. In the case of \ce{H2O} pure rotational lines the model overpredicts the ratio, which is found to be marginally lower than unity in the observed spectrum. Numerical experiments indicate that such observed behavior can be replicated by fine-tuning the FUV part of the stellar spectra. In fact, \cite{Vlasblom2025} show that a lower FUV luminosity from the central star can increase the abundance of \ce{H2O} in the upper disk layers, and in consequence, the strength of the cold water emission. Additionally, recent thermochemical modeling by \cite{Calahan2026} show that the strength of the correlation between the observable mass of different water reservoirs and accretion luminosity depends on the location of each reservoir relative to the optically thick dust surface. The location of that surface depends on different properties of the dust population, including the level of settling and gas-to-dust ratio. We recommend for future campaigns to include dedicated FUV observations to understand the effect of a variable stellar spectra on the reservoirs of hot- and cold-water in protoplanetary disks. Finally, we note that a higher signal-to-noise ratio is required to confirm the predicted line intensifications for \ce{CO2} and \ce{HCN}.      
       
\subsection{Disk Structure Diagnosis}
\label{subsec:convergedStruct}
\begin{figure}
\includegraphics[width=\hsize]{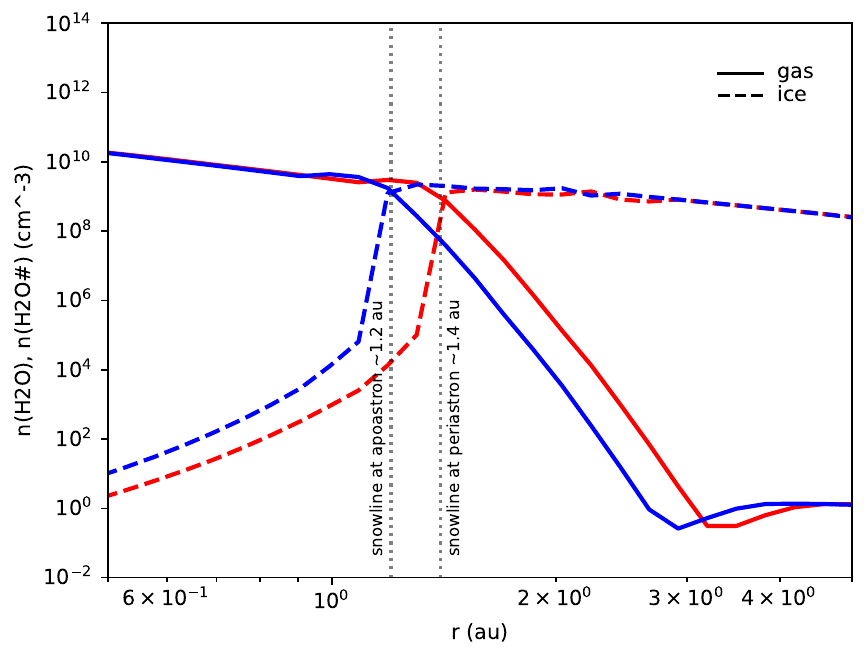}
\caption{\rev{Gas- and solid-phase water abundances in the disk midplane at periastron (red) and apastron (blue)}. Results are shown for two disk models: periastron (including the accretion burst; red) and apastron (without the accretion burst; blue). Both models employ the "nf" X-ray configuration, accounting only for baseline X-ray emission. The water snowlines indicate where the gas- and solid-phase abundances are equal; these locations are marked by the vertical dotted lines.}
\label{fig:snowline}
\end{figure}

The snowline location serves as a meaningful tracer of disk structure changes induced by variable stellar irradiation. We define the snowline as the radius at which the gas-phase water abundance equals the ice-phase abundance (Fig. \ref{fig:snowline}). We find that at periastron the snowline shifts outward by 17\% relative to its apastron position. This suggests that, although the effect is less dramatic than in outburst systems—where the snowline can shift outward by up to 100\% compared to its quiescent location \citep{Smith2025,Rab2017} --- even routine accretion variability can induce temporal changes in the snowline position.  

The predicted shift in the snowline is consistent with the expectation for regions in the disk that are optically thin to thermal radiation. In fact, in the optically thin limit, $R_\mathrm{periastron}^\mathrm{snow}/R_\mathrm{apastron}^\mathrm{snow} \sim (L_\mathrm{periastron}^\mathrm{tot}/L_\mathrm{apastron}^\mathrm{tot})^{1/2}$. By integrating the profiles in Fig. \ref{fig:starSpec}, we find  $R_\mathrm{periastron}^\mathrm{snow}/R_\mathrm{apastron}^\mathrm{snow} \approx 1.18$, which is in agreement with the $17\%$ shift predicted by the numerical simulations. However, we note that shifts in the snowline in the optically thick regions may be modulated by time-dependent radiation diffusion (see Sect. \ref{subsec:model_caveats}), similar to what was modeled for the more energetic FU Ori outbursts \citep{Laznevoi2025}.

\begin{figure}
\includegraphics[width=\hsize]{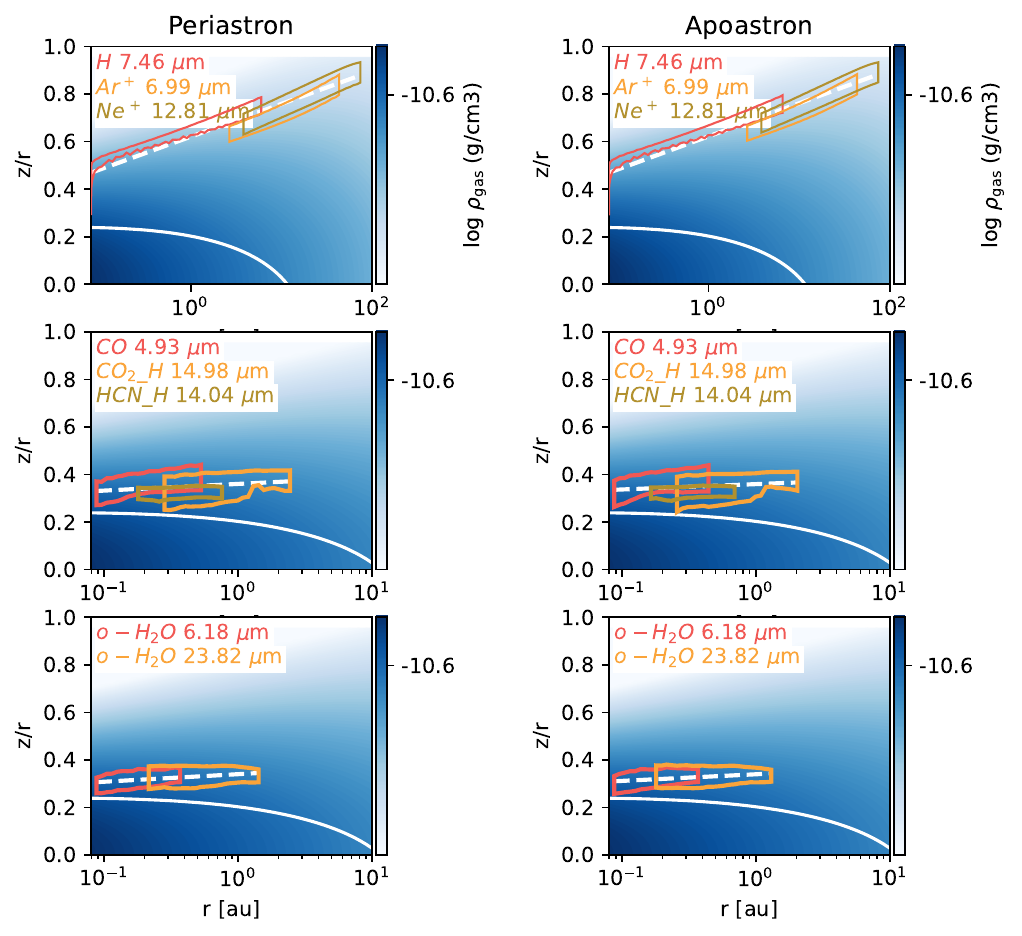}
    \caption{Two-dimensional emitting regions for atomic and molecular species derived from the same thermochemical models as in Fig. \ref{fig:snowline}. Colored polygons enclose the 15\%--85\% emitting areas for the transitions indicated in the legends. The white solid lines denote the $\tau_V=1$ vertical surface, while the white dashed lines indicate the representative surfaces of emission for the respective species. The value indicated in the color bar corresponds to a gas density equal to $1/e$ times the value in the midplane at $1$ au.}
\label{fig:emittingAreas}
\end{figure}

Additionally, we use the models to inform about the behavior of the emitting regions for those species detected with JWST/MIRI. Figure~\ref{fig:emittingAreas} displays the emitting areas\footnote{The emitting area of a line is defined \rev{in terms of the 15\% and 85\% quantiles of the flux spatial distribution. The left and right edges of the emitting areas correspond to the radii where the cumulative flux reaches 15\% and 85\% of the total flux emitted from the disk. At each radius, the lower and upper edges indicate the heights enclosing 15\% and 85\% of the flux emitted within the corresponding vertical column, measured from the disk’s atmosphere down to the midplane}. For this analysis, we ignore the interplay between dust and gas source functions when calculating line photon escape probabilities.} for the transitions indicated in the figure legends for H I (6-5), \ce{[Ar II]}, and \ce{[Ne II]} (colored polygons, upper panels); \ce{CO}, \ce{CO2}, and \ce{HCN} (middle panels); and \ce{H2O} near 6~$\mu$m and 24~$\mu$m (lower panels), all overlaid on gas density maps. Our results show a vertical stratification of the line emission: an upper layer containing the atomic lines, a middle layer dominated by \ce{CO}, \ce{CO2}, and \ce{HCN} emission, and a lower layer, located slightly deeper, from which \ce{H2O} emission originates. To characterize these regions, we performed polynomial fits to the vertical mid-points of the emitting areas measured at each radial distance. The resulting surfaces --- which we refer to as "representative surfaces of emission" --- are indicated by dashed lines in Fig.~\ref{fig:emittingAreas}.
      
\begin{figure}
\includegraphics[width=\hsize]{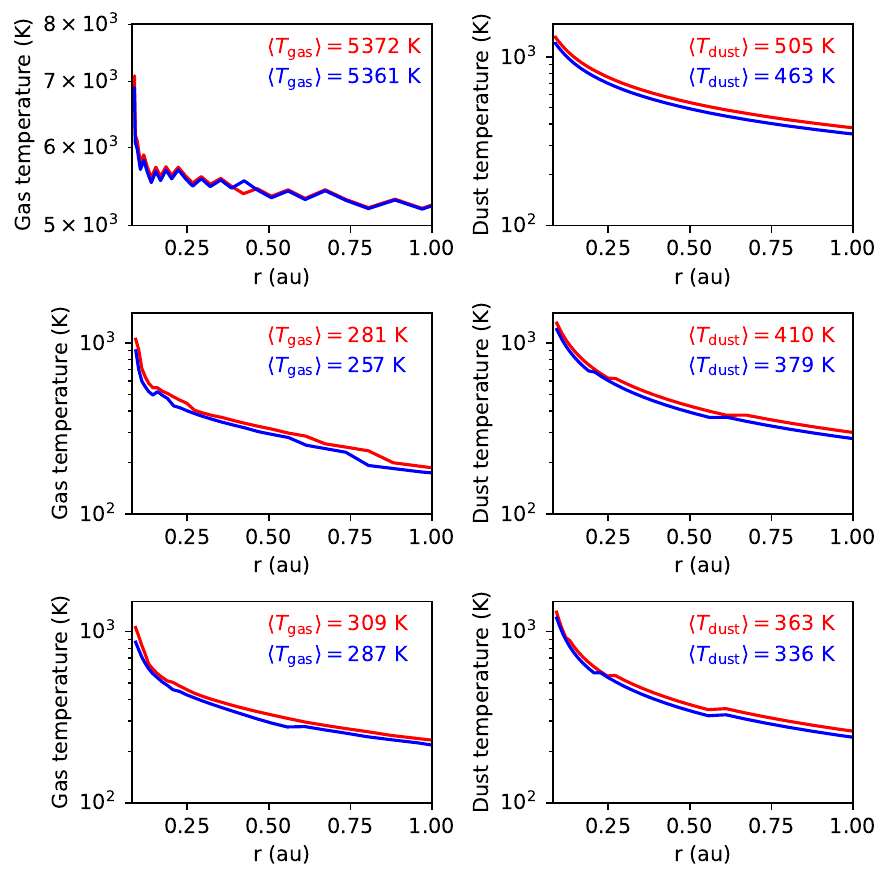}
\caption{Gas and dust temperature profiles extracted along the characteristic emission surfaces for \ce{H I}, \ce{[Ar II]} and \ce{[Ne II]} (upper panels)}; \ce{CO}, \ce{CO2}, and \ce{HCN} (middle panels); and \ce{H2O} (lower panels). These results are derived from the same periastron (red) and apastron (blue) models used in Figs. \ref{fig:snowline} and \ref{fig:emittingAreas}. (\textit{left:})
\label{fig:diagnosis}
\end{figure}

We subsequently retrieved the gas and dust temperatures along each representative surface for both orbital phases and calculated their radial averages\footnote{The average temperature is defined as $\langle T \rangle = \dfrac{\int T(r) \cdot r \, dr}{\int r \, dr}$.} within the first astronomical unit. As illustrated in Fig. \ref{fig:diagnosis}, gas and dust temperatures at periastron are systematically higher than those at apastron across the emitting surfaces. This is more notorious in the middle and lower layers, where the gas is 10\% warmer at periastron than at apastron. The gas temperature in the uppermost layer, where the leading heating mechanism is X-ray coulomb heating, is essentially the same between epochs. The presence of an X-ray flare at periastron, however, raises the gas temperature of this uppermost layer by 4\% (see Sect. \ref{subsec:isoXrays} and Fig. \ref{fig:diagnosis-flare}). Dust temperature increases by 10\% between orbital phases. However, along the middle and lower representative surfaces, the dust temperatures exceed the gas temperatures. This result is unexpected as for realistic, vertically non-isothermal disks, we expect $T_{\rm{gas}}(z) \geq T_{\rm{dust}}(z)$ for all $z$. An improved forward modeling for the DQ Tau disk that combines both dust and gas constraints is needed to solve this issue, but such an effort is beyond the scope of this paper.

Finally, we explored the effect of an enhanced accretion luminosity on the vertical structure of the disk. We simulate the disk in hydrostatic equilibrium mode, where the vertical distribution of gas is determined by a balance between gas pressure and stellar gravity. The vertical structure can be conveniently characterized by the vertical scale height $h = c_s/\Omega$; i.e., the ratio of the sound speed to the keplerian frequency of a gas parcel, both measured at the midplane. We extract each quantity from the model output and find that, in the inner disk, gas is only $\sim 5\%$ more vertically extended at periastron than at apastron.

\subsection{Steady State Molecular and Atomic Column Densities}
\label{subsec:molAbun}
\begin{figure*}
\includegraphics[width=\hsize]{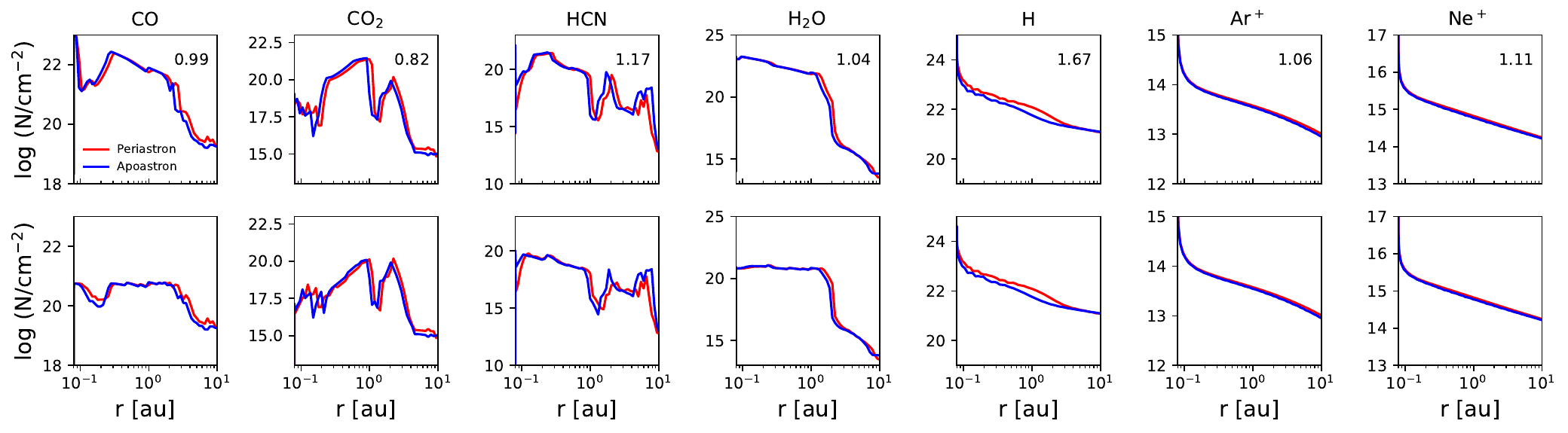}
\caption{Column density profiles for atomic and molecular species derived from the same periastron (red) and apastron (blue) thermochemical models used in Figs. \ref{fig:snowline}--\ref{fig:diagnosis}. Profiles are integrated over the total vertical extent of the disk (\textit{top panels}) and from the disk surface down to the $\tau_V=1$ dust photosphere (\textit{bottom panels}). Legends indicate the mass ratios (periastron/apastron) for each species, calculated by integrating the surface densities from the inner disk rim out to $10$~au.}
\label{fig:abundances}
\end{figure*}

We do not aim to an in-depth point-by-point chemistry analysis, and rather focus on interpreting trends that are systematic. Steady state chemical abundances are vertically integrated and their radial profiles are displayed in Fig. \ref{fig:abundances}. Upper panels depict the total column density; this is, the integral is evaluated from the disk's uppermost layer down to the midplane. By integrating the profiles over the radial and azimuthal coordinates, we compute the mass of each molecule (in the gas phase) contained within the first $10$ au. The periastron-to-apastron mass ratio is indicated in the upper-right corner of each panel in Fig. \ref{fig:abundances}. Lower panels display the fraction of the column that is visible above the $\tau_\mathrm{V}=1$ vertical surface. 

The radial column-density profiles of the atomic species exhibit only minor variations between periastron and apastron. The most noticeable difference is found for hydrogen, whose abundance is systematically higher during periastron within the inner 10 au. This enhancement results from an increased \ce{H2} dissociation rate in the uppermost layers of the disk. The abundances of \ce{Ar+} and \ce{Ne+} remain largely unchanged between orbital phases. However, as shown in Fig. \ref{fig:abundances-flare}, the ionizing effect of a strong X-ray flare significantly enhances the abundances of these ions, increasing them by approximately a factor of two in the innermost regions of the disk, consistent with previous findings (e.g., \citealt{Aresu2011}).      

The abundances of molecular species are also only marginally affected. Trends are less pronounced as the radial profiles at the two epochs intersect at several radial locations. For water, the most notable feature is the shift in the radial profile during periastron, which is driven by the displacement of the water snowline discussed in Sect. \ref{subsec:convergedStruct}

\section{Discussion} 
\label{sec:discussion}
\subsection{Investigating the Effect of an X-ray Flare}
\label{subsec:isoXrays}
As noted in Sect. \ref{subsec:chandra_and_swift_observations}, our \textit{JWST} observations were not contemporaneous with the X-ray flares detected by \textit{Chandra} and \textit{Swift}. However, during the Jan. 28--31 and Mar. 1--4, 2025 periastron passages, two flares were captured with peak luminosities $L_X \gtrsim 10^{31}$~erg~s$^{-1}$ (Fig. \ref{fig:chandra_swift_dqtau_2025}). It is therefore instructive to model the response of the mid-IR spectrum to an X-ray flare of similar magnitude. To this end, we performed a third simulation (variant ``f'') using a stellar spectrum with a periastron X-ray component of $1.09 \times 10^{31}$~erg~s$^{-1}$ and a plasma temperature of $54$~MK. The resulting synthetic flux ratios are presented in Fig. \ref{fig:fluxRatios} and in Table \ref{tab:wavelength-ranges} (``f" column).

For the atomic lines, the periastron-to-apastron ratios are notably enhanced compared to the non-flaring case. Because the apastron baseline remains constant across models, these discrepancies are driven solely by the X-ray flare. To identify the physical drivers of this luminosity boost, \rev{we perform a detailed point-by-point chemical analysis of the flare solution. For this analysis, we focus on the grid cell with coordinates $(r,z)=(1.0,0.64)$ au, which lies within the emitting region of the \ce{H (6-5)} line.} 

\rev{As expected, we find that the main creation and destruction channels of \ce{H,\ Ar+} and \ce{Ne+} are driven by X-rays. For neutral hydrogen, the main creation route is the radiative recombination reaction \ce{H+ + e- \rightarrow H + \nu}, with a formation rate of $6 \ \rm{cm}^{-3}\ \rm{s}^{-1}$ at point $(r,z)$. This channel is two orders of magnitude more efficient than the second leading creation route, the charge exchange reaction \ce{H+ + O \rightarrow O+ + H}, which shows a formation rate of only $3 \times 10^{-2} \ \rm{cm}^{-3}\ \rm{s}^{-1}$ at the same disk location. The same pattern is seen for \ce{Ar+} and \ce{Ne+}; this is, the main formation route for the species in the $n$-th ionization state is the radiative recombination of its $(n+1)$-th ionization state. Evidently, stronger X-ray and EUV radiation fields significantly increase the ionization rate throughout the inner disk, which in turn enhances the abundances of \ce{H+,\ Ar++} and \ce{Ne++}. In addition, X-ray photons can trigger single and double ionization when absorbed in the K-shells of abundant elements such as \ce{He,\ C,\ N} and \ce{O}. The energetic electrons generated in these events can further ionize the surrounding gas via secondary ionization, a phenomenon that is accounted for in our models \citep{Aresu2011}. In conclusion, the enhanced abundances of elements in higher ionization states and free electrons during a flare lead to a higher number of recombination events, which explains the higher luminosities of the atomic lines predicted by our models.} Figure \ref{fig:abundances-flare} shows the radial column densities for both the molecular and atomic species. By integrating the radial profiles of the atomic species, we find that their mass reservoirs increase by a factor of two compared to apastron values. This stands in stark contrast to the non-flaring case (Fig. \ref{fig:abundances}), where the abundances of \ce{H}, \ce{Ar+}, and \ce{Ne+} remain nearly constant between epochs.

These results are particularly relevant for hydrogen recombination lines. Previous studies \citep{Rigliaco2015, Tofflemire2025,Shridharan2025} identified a positive correlation between accretion luminosity and the luminosities of the \ce{H I} (7--6), (10--7), and (8--7) lines, attributing their origin to accretion columns. Our models support this interpretation: in both the non-flaring and flaring cases, the disk-only emission significantly underestimates the observed flux, implying a non-disk component is required to match the data. Furthermore, we show that X-ray flares can appreciably enhance the luminosity of these hydrogen lines. We therefore advocate for coordinated multi-wavelength campaigns---simultaneously utilizing \textit{Chandra}, \textit{Swift}, \textit{JWST}, and ground-based optical facilities---to observationally assess the effect of X-rays on the luminosity of the mid-IR recombination lines. Given the frequency of these X-ray flares, this will require more than a single \textit{JWST}/MIRI snapshot per periastron \citep[as obtained by][]{Kospal2025} to effectively cover longer periastron spans (Fig. \ref{fig:chandra_swift_dqtau_2025}). 

Finally, our models indicate the X-ray flare has a negligible effect on both the abundances and line luminosities of the molecular species detected in DQ Tau: \ce{CO}, \ce{CO2}, \ce{HCN}, and \ce{H2O}.

\subsection{Model Caveats}
\label{subsec:model_caveats}
Our calculation of the gas temperature assumes equilibrium between the rate of internal energy flowing into a volume element and its rate of energy loss. The time evolution of the internal energy density of a volume element is:

\begin{equation}
\label{eq:thermal_balance}
    \frac{de}{dt}=\sum_i \Gamma_i - \sum_j \Lambda_j = f(e),
\end{equation}

\noindent where $\Gamma_i$ and $\Lambda_j$ are the heating and cooling functions summed over all the possible heating and cooling mechanisms considered in the model. Therefore, our thermodynamic equilibrium assumption implies the existence of an equilibrium energy density, $e_0$, that balances out heating and cooling.

Let us assume a small perturbation around equilibrium of the form $e=e_0+\delta e$. By performing a linearization of Eq. \ref{eq:thermal_balance} around $e_0$ and retaining only linear terms in $\delta e$, we get $\delta e \propto e^{-t/\taurel}$, where the relaxation timescale is defined as

\begin{equation}
\label{eq:relaxation_timescale}
    \taurel = \bigg|\,\frac{df}{de} \Big |_{e_0}\, \bigg|^{-1}.
\end{equation}

\noindent It is evident from Eqs. \ref{eq:thermal_balance} and \ref{eq:relaxation_timescale} that the relaxation timescale strongly depends on the behavior of the heating and cooling functions. In our models, those functions are spatially dependent, since different heating and cooling mechanisms dominate at different locations in the disk. We corroborate this by examining the model output. We find that the relaxation timescale along the representative emitting surfaces ranges from a fraction of a day near the inner rim to approximately one hundred days at an orbital distance of $\sim 1$ au. Since those timescales are comparable to both the orbital period of the binary ($\sim 16$ days) \rev{and typical X-ray flare light curves ($\sim$ a few hours)}, a comprehensive study of DQ Tau's thermochemical response to temporal variations in the binary's radiation field must consider non-equilibrium thermodynamics.

A few works have addressed the problem of time-dependent radiative transfer in circumstellar disks. \cite{Bensberg2022} extended the \texttt{POLARIS} code \citep{Reissl2016,Reissl2018} and implemented a time-dependent algorithm to solve for the dust temperature in an optically thin disk subjected to a fourfold enhancement of the stellar luminosity due to an outburst. They found that, while the post-outburst heatwave travels across the $100$ au disk in roughly $1$ day, the dust grains relax to equilibrium in just a few seconds after being perturbed by the wave. Thus, our equilibrium solution for the dust temperature seems well justified, at least in the optically thin regions of DQ Tau. Similarly, \cite{Pavlyuchenkov2024} introduced a method to calculate the non-stationary thermal structure of a disk based on splitting the radiation field into a stellar and a disk component. Similar to our case, the author finds that the thermal relaxation timescale varies across the disk, with timescales under $0.1$ yr for distances below $1$ au.  \cite{Harries2011} investigated the response of a disk’s SED to periodic accretion variability and found a wavelength-dependent time lag between the NUV/optical emission due to accretion shocks and the reprocessed infrared emission from the disk. For $\lambda \sim 10\ \micron$, the author inferred a time lag of only $\sim 500$ seconds, which is notoriously lower than our relaxation timescales. The discrepancy is likely explained by the different approaches to solve the heating-cooling balance: \cite{Harries2011} assumes a gas component in LTE where black body cooling is counterbalanced by radiative heating. In contrast, our models consider $119$ heating processes and $111$ cooling processes in a feedback loop with a large chemical network and independent gas and dust temperature calculations. 

Non-equilibrium chemistry considerations are equally important. \cite{Ilgner2006} studied the effect of a periodic sequence of X-ray outbursts on the structure of a disk's dead zone and found that under favorable conditions of reduced opacity, enhanced plasma temperature and presence of heavy metals in the disk, the structure of the dead zone can in fact be modulated by the X-ray outburst frequency. \cite{Waggoner2019} present models of time-dependent water chemistry in disks subjected to evolving X-ray flaring events. Their models suggest water abundance increments that are followed by abundance decays \rev{on} timescales that are either shorter or longer than $10$ days, depending on the location in the disk. The authors also studied the response of water abundance to X-ray flares of different strengths. In particular, for X-ray luminosities two to five times stronger than a nominal value of $L_\mathrm{X}=4\times 10^{30} \ \mathrm{erg}\ \mathrm{s}^{-1}$, they predict water abundance enhancements lower than a factor of two, which is consistent with our findings. In a subsequent publication, \cite{Waggoner2022} examine the effect of the stochastic nature of X-ray flaring events on time-dependent disk chemistry and found that chemical changes occurring \rev{on} timescales of a few days are due to discrete flaring events, whereas changes occurring \rev{on} timescales of centuries are due to the cumulative effect of many individual flares. Additionaly, their simulations show that cations are particularly sensitive to flares, whereas other species such as \ce{HCN} and \ce{CO} change their abundances by less than $1\%$. Finally, it is worth mentioning that the interplay between dynamical processes in the disk, such as turbulent diffusion, and chemical evolution is not treated in detail in our models (e.g., \citealt{Semenov2011}).

We highlight that none of the works described above have self-consistently modeled the complex interplay among time-dependent radiative transfer, heating-cooling balance, and chemistry. Such an effort would require a complete (magneto-)hydrodynamical treatment coupled to a robust solution of the disk chemistry. To our knowledge, such a self-consistent modeling framework is not currently available. Nevertheless, our results indicate that a time-independent treatment is already capable of explaining reasonably well the observed behavior of the most important gas tracers in the mid-IR.

\subsection{Implications for the Observed Spitzer/JWST Line Variability}
\label{subsec:implicationForSpitzerJWST}
At the time of writing, only four disks observed with both \spitzer\ and JWST/MIRI have their $\sim 15$ year baseline line spectra presented and discussed in the literature: SZ Cha \citep{Espaillat2023}, EX Lup \citep{Kospal2023,Smith2025}, AS 209 \citep{Romero-Mirza2024}, and VW Cha A \citep{Kurtovic2026}. The most extreme case of line variability is seen in the EX Lup disk, where observations during the 2008 outburst \citep{Banzatti2012} showed a strong enhancement in \ce{H2O} and \ce{OH} fluxes, and the disappearance of the \ce{C2H2}, \ce{HCN} and \ce{CO2} signal. This contrasts with the most recent JWST observations of EX Lup in quiescent state \citep{Kospal2023}, where the signal from the organics is recovered, and an enhanced cold water component is detected \citep{Smith2025}. 
 
Our results imply that the line emission variability observed in non-outbursting systems–––where flux ratios of roughly a factor of two are seen in the mid-infrared–––is consistent with, and can be explained by episodes of routine accretion variability. We showed this for the case of DQ Tau, whose circumbinary configuration allows for predictable, multiwavelength monitoring of accretion events at different orbital phases. For disks around single stars this is more challenging, but one can estimate the probability of observing such a system during an accretion burst. In fact, the repeat timescales of bursting events for disks across different star-forming-regions has been measured to range from $\sim 3$ to $80$ days \citep{Cody2017} which translates into a probability of approximately $24\%$ down to $1\%$ of observing a disk in a bursting event on any given day, respectively. Therefore, observing a disk during an accretion burst is not an unusual event for repeat timescales shorter than $19$ days.

Our results also suggest that X-ray flares of moderate intensity have a negligible effect on the mid-IR molecular emission. However, X-rays have a prominent effect on the atomic lines. We emphasize that the synthetic fluxes presented in this work originate in the upper layers of the disk, not in accretion columns. Our results imply that, if an accretion event is simultaneous with an X-ray flare, caution must be exercised when using recombination lines to derive accretion luminosities, as the line luminosity may include a non-negligible contribution from the disk. However, such a scenario is rather unlikely. In fact, assuming an occurrence rate of one $E_\mathrm{X} \sim 10^{34} \mathrm{erg}$ flare every two days \citep{Getman2021}, the probability\footnote{Assuming accretion bursts and X-ray flares are statistically independent events.} of observing a system in both a bursting \textit{and} a flaring state on any given day  ranges from $0.3\%$ (for a burst timescale of $80$ days) to $7\%$ (for a burst timescale of $3$ days).

\section{Conclusions} 
\label{sec:conclusions}
We investigate the impact of routine accretion variability and moderate-intensity X-ray flares on the thermochemical structure of a protoplanetary disk. Leveraging contemporaneous multi-wavelength observations of the DQ Tau system---spanning X-ray to mid-infrared (mid-IR) regimes---we utilize thermochemical modeling to assess the observational signatures of these events in the mid-IR line spectra, as probed with JWST/MIRI \citep{Kospal2025}, during consecutive periastron and apastron passages. Our main conclusions are summarized below: 

\begin{enumerate}
    \item We obtained multi-epoch observations of DQ Tau using \textit{Chandra}/ACIS-I and \textit{Swift}/XRT-UVOT during the periastron passages of January 28-31, February 13-16, and March 1-4, 2025. The intervening apastron passage of February 7-9, 2025, was also monitored with \textit{Chandra}/ACIS-I. From these data, we derive a baseline X-ray luminosity of $2 \times 10^{30}$~erg~s$^{-1}$. Two high-amplitude X-ray flares were detected during the January and March windows, reaching characteristic luminosities of $\sim 1.1 \times 10^{31}$~erg~s$^{-1}$. We note, however, that neither flare was contemporaneous with our JWST/MIRI observations (Sect.~\ref{subsec:chandra_and_swift_observations}). Both new and archived \textit{Chandra} and \textit{Swift} data were utilized to construct the X-ray/EUV stellar spectral energy distributions (SEDs) used as inputs for our thermochemical models. The new \textit{Swift} data --- collected during the January 28-31, 2025 periastron (peak accretion) and the onset of the February 13-16, 2025 periastron (global accretion minimum) --- revealed a fivefold increase in the \texttt{M2} ($\sim 225$~nm) and \texttt{W1} ($\sim 260$~nm) fluxes (Sect. \ref{subsec:XraySpec}, \ref{subsec:UVSpec}).\\  

    \item We employ thermochemical models illustrative of the DQ Tau disk at both orbital phases --- specifically the "nf" (no-flare) case, which excludes X-ray flaring --- to evaluate the impact of the variable stellar spectrum on the disk's physical structure and mid-IR line emission. The wavelength-integrated line fluxes from \ce{CO}, \ce{CO2}, \ce{HCN}, and \ce{H2O} are systematically stronger at periastron than at apastron, responding to the increase in accretion luminosity from $0.1~L_\odot$ to $0.4~L_\odot$. The \ce{H I}, [\ce{Ar II}] and [\ce{Ne II}] lines originating in the disk atmosphere remain largely unaffected by the accretion burst (Sect. \ref{subsec:effectOnSpec}). \\ 

    \item Using our baseline thermochemical models, we derive periastron-to-apastron line flux ratios ranging from $1.1$ to $1.6$ across the mid-IR spectrum. The synthetic ratios for \ce{CO, \ CO2, \ HCN} and the rovibrational \ce{H2O} lines are in good agreement with the observations (Sect. \ref{subsec:comparetoDQTau}). The observed pure rotational \ce{H2O} emission around $24~\mu\mathrm{m}$ exhibits the opposite trend and is marginally stronger at apastron than at periastron, a behavior that our models fail to replicate. Contemporaneous FUV constraints to the stellar spectra are required to interpret the origin of such differential response of the different water reservoirs. Furthermore, the predicted ratios for [\ce{Ar II}] and [\ce{Ne II}] align well with the data. However, our models fail to reproduce the observed ratios of the H~I (6--5), (10--7), (7--6), and (8--7) recombination lines. This discrepancy supports the interpretation that these lines originate within accretion columns, as established in previous studies.\\

    \item Routine accretion variability produces only modest changes in the thermochemical structure of a disk. Taking the thermochemical solution at apastron as a reference, we find that at periastron: the disk vertical scale height increases by $5\%$; the snowline recedes outwards by only $\sim 17\%$ in the midplane; gas and dust temperatures increase by approximately $10 \%$; and the steady-state mass reservoirs of atomic and molecular species in the inner disk remain mostly unaffected (Sect. \ref{subsec:convergedStruct} and \ref{subsec:molAbun}).\\  

    \item Our thermochemical models --- specifically the ``f'' variant, which incorporates a periastron X-ray flare ---  suggest that X-ray flares of moderate peak luminosity ($\sim 1.1 \times 10^{31}$~erg~s$^{-1}$; i.e., five times the baseline level for DQ Tau) do not significantly alter the mid-IR molecular emission. Such flares do, however, notoriously increase the chemical abundances and line strengths of \ce{H I}, \ce{Ar II}, and \ce{Ne II} in the disk surface (Sect. \ref{subsec:isoXrays}).
\end{enumerate}

Our results imply that the \spitzer/JWST mid-IR line variability observed in non-outbursting systems could be explained by episodes of routine accretion variability and X-ray flaring events in the host star. While other processes might also contribute, a time-varying stellar spectrum with moderate intensity fluctuations---as expected for most pre-main-sequence stars---provides a simpler alternative explanation for the observed variability.

\newpage


\newpage

\begin{acknowledgements}
      \rev{We thank Dr. Manuel Güdel for an insightful and constructive referee report which enhanced the quality of the manuscript.} This work is supported by STScI grant \#JWST-GO-04876.002-A and SAO \textit{Chandra} grant GO5-26005X. This work is based on observations made with the NASA/ESA/CSA \textit{James Webb Space Telescope}. The data are associated with programme 4876 and were obtained from the Mikulski Archive for Space Telescopes (MAST) at the Space Telescope Science Institute, which is operated by the Association of Universities for Research in Astronomy, Inc., under NASA contract NAS 5-03127. This research has made use of data obtained from the \textit{Chandra} Data Archive and software provided by the \textit{Chandra} X-ray Center (CXC), which is operated by the Smithsonian Astrophysical Observatory for and on behalf of NASA under contract NAS8-03060. We also acknowledge the use of data obtained from the \textit{Swift} data archive. \rev{The Center for Exoplanets and Habitable Worlds is supported by the Pennsylvania State University and the Eberly College of Science.} This work was supported by the ADVANCED 149943 grant, which has been implemented with the support provided by the Ministry of Culture and Innovation of Hungary from the National Research, Development and Innovation Fund, financed under the NKKP ADVANCED funding scheme. D.~S. was funded by the Deutsche Forschungsgemeinschaft (DFG, German Research Foundation) – project number: 550639632. 
\end{acknowledgements}

%

\bibliographystyle{aa}
\bibliography{references}







   
  



\begin{appendix}




\onecolumn
\section{}
This Appendix presents radial profiles of column densities and gas/dust temperatures along characteristic emission surfaces, comparing the baseline apastron model (blue) with a periastron model that incorporates both the accretion burst and the large X-ray flare (violet).

\begin{figure}[H]
\includegraphics[width=\hsize]{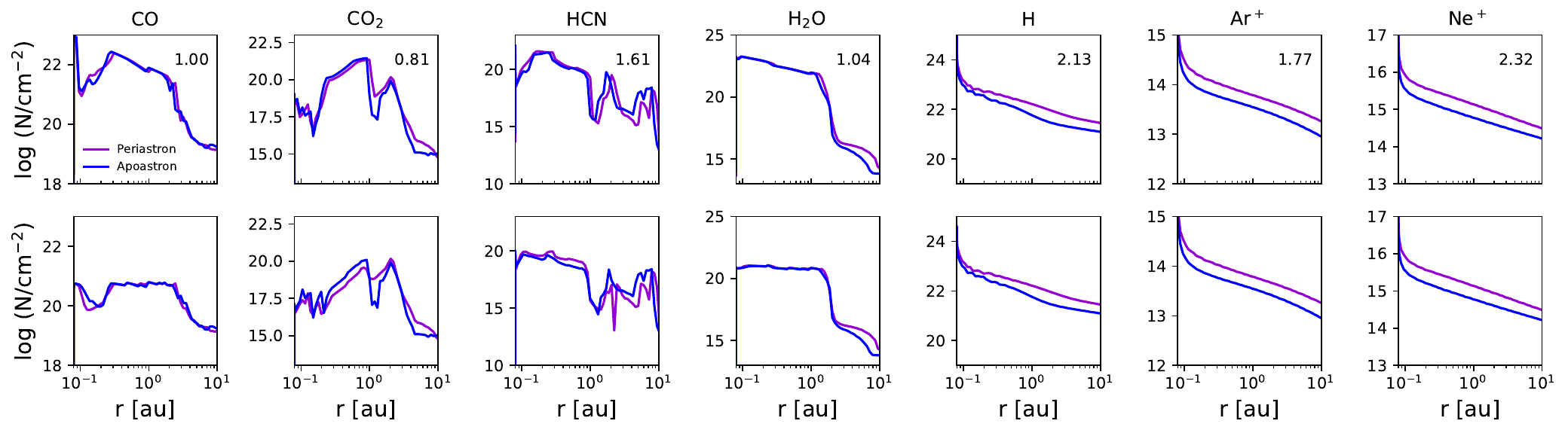}
\caption{Same as Fig. \ref{fig:abundances}, but comparing the baseline apastron model (blue) with a periastron model (violet) that incorporates both the accretion burst and the large X-ray flare. \label{fig:abundances-flare}}
\end{figure}

\begin{figure}[H]
\centering
\includegraphics[width=0.5\hsize]{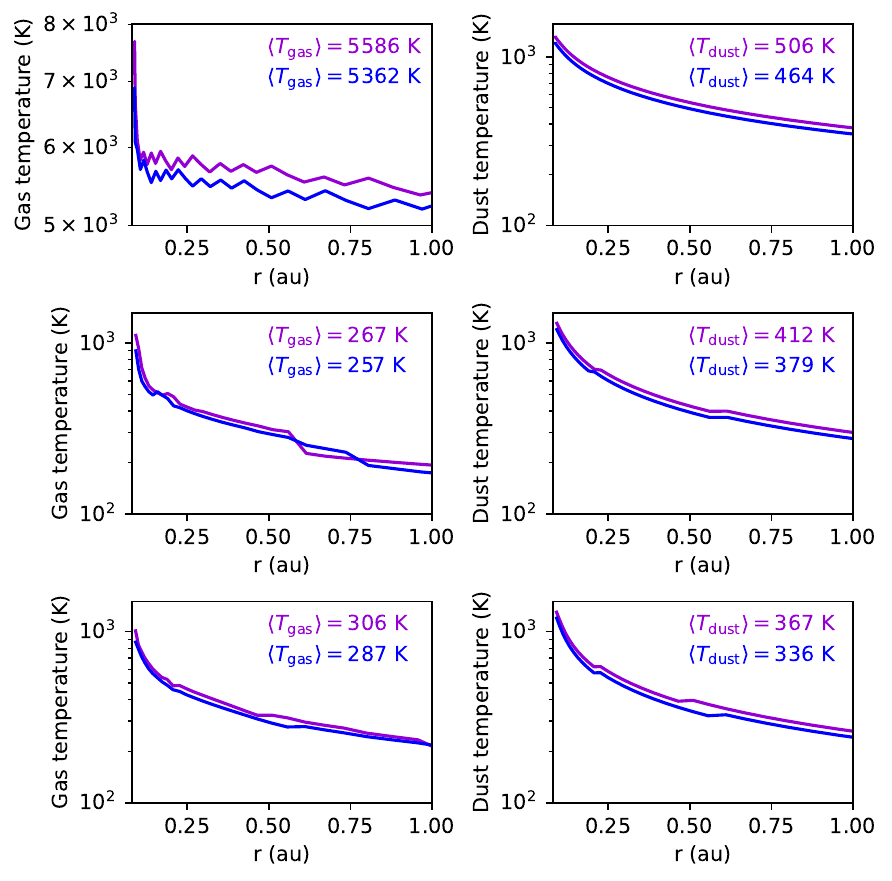}
\caption{Same as Fig. \ref{fig:diagnosis},  but comparing the baseline apastron model (blue) with a periastron model (violet) that incorporates both the accretion burst and the large X-ray flare. \label{fig:diagnosis-flare}}
\end{figure}

\section{The Effect of Dust Settling}
\label{subsec:effect-of-settling}
To explore the effect of a vertically-stratified distribution of grains on the line ratios, we run Models 1 and 2 (see Sect. \ref{subsec:effectOnSpec}), this time following \cite{Dubrulle1995} prescription for dust settling. We adopt a \cite{Shakura1973} $\alpha$ factor of $10^{-3}$. 

\begin{figure}[H]
\includegraphics[width=\hsize]{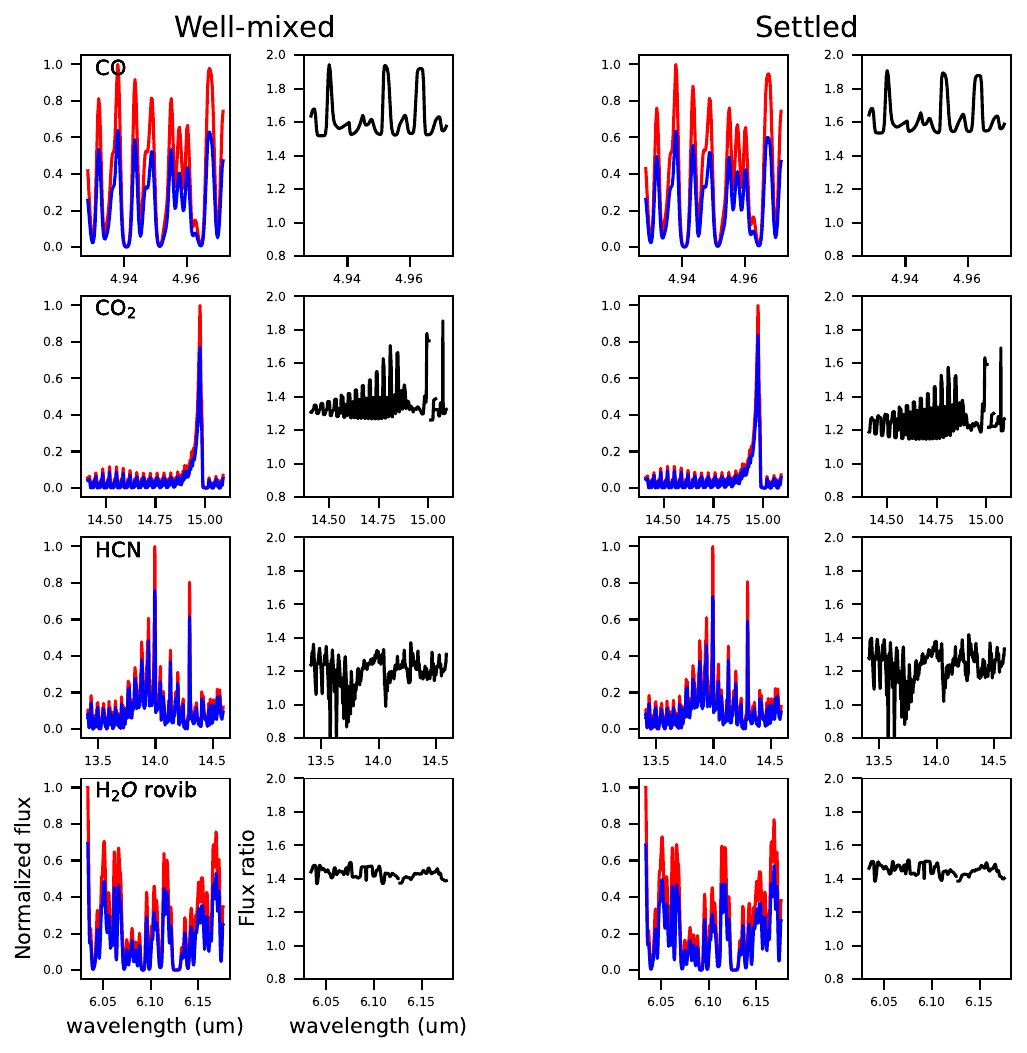}
\caption{The effect of dust settling on the periastron-to-apastron line fluxes. The first and second columns show the results for the well‑mixed case, while the third and fourth columns present the settled case. Each row corresponds to a different species. For each case, the first column displays the normalized spectra at periastron (red) and apastron (blue), and the second column shows the periastron-to-apastron flux ratio computed at each wavelength.}
\label{fig:settling-effect}
\end{figure}

As expected, the fluxes at periastron and apastron differ in absolute scale between the well-mixed and settled cases. However, the line ratios from one epoch to another are essentially the same for both cases. Figure \ref{fig:settling-effect} shows the spectra of the molecular species retrieved from the well-mixed and settled simulations. For each case, left columns display the spectra normalized to the maximum flux at the respective periastron. Right columns display the periastron-to-apastron line ratio computed at each wavelength.

The line ratios are similar between the well‑mixed and settled cases because settling leads to differences in excitation conditions that are comparable to those found in the well-mixed case. Therefore, our conclusions based on line ratios remain robust to the effects of dust settling.

\end{appendix}
\end{document}